\documentclass[a4paper,11pt]{article}
\usepackage{pos}

\usepackage[sort&compress,numbers]{natbib}
\title{Probing ultralight and degenerate dark matter with galactic dynamics}

\author*[a,b]{Diego Blas}

\affiliation[a]{Institut de Fisica d’Altes Energies (IFAE), The Barcelona Institute of Science and Technology, Campus UAB, 08193 Bellaterra (Barcelona), Spain}

\affiliation[b]{Instituci\'o Catalana de Recerca i Estudis Avan\c cats (ICREA), Passeig Llu\'is Companys 23, 08010 Barcelona, Spain}

\emailAdd{dblas@ifae.es}

\abstract{This short contributions summarizes a couple of recent results to test dark matter properties with galactic dynamics. First, I will present the impact in rotation curves from solitonic structures expected at the center of galaxies for  ultralight bosonic dark matter. As a result, one can claim that masses of the order 
$m_{\rm DM}\lesssim 10^{-21}$eV are in tension with data. Second, I will discuss how the dark matter medium properties change the way a `probe' interacts with the halo. I will focus on dynamical friction and show how it is modified in the case of degenerate fermions. This result may be used to address the Fornax timing problem.
I hope that this contribution represents an inspiration to continue exploring other ideas in this direction of using galactic dynamics to tell apart different dark matter models.}

\FullConference{1st General Meeting and 1st Training School of the COST Action COSMIC WSIPers (COSMICWISPers)\\ 
5-14 September, 2023\\
Bari and Lecce, Italy\\}

\begin{document}
\maketitle

\section{Introduction}

\vspace{-.3cm}
The existence of a dark matter (DM) component plays a key role in the properties of visible matter in galaxies. One of the most significant is the modification of the gravitational potential felt by the visible matter component, which eventually modifies properties such as the rotation curves of visible matter at galactic scales \cite{Bertone:2010zza}. However, when one considers  dark matter as a \emph{medium} in which the visible matter is moving, several new dynamical effects may be present, e.g. tidal disruption, dynamical friction, dynamical heating \cite{Binney,2010gfe..book.....M}. These effects have been clarified and a significant amount of dark matter phenomenology is known for the models where dark matter at galactic scales can be understood as a collection of particles of small mass and only interacting gravitationally. However, these \emph{medium} properties may change if dark matter is made of compact objects, ultralight DM (ULDM) or if other forces act in the dark sector. As a result, these other proper properties of DM can be used to differentiate among dark matter models, and may be even enhanced in some cases. A long-term goal  suggest to study them for the different dark matter models, and use \emph{galactic dynamics for fundamental physics}, inspired by the related efforts related to stars  \cite{Raffelt:1999tx}.

In this short contribution, I will focus on two particular directions I have worked on. First, I will discuss how
\emph{galactic rotation curves} can be used to test ultra-light bosonic\footnote{Due to lack of time, I won't discuss in detail similar bounds on the fermionic DM case, see e.g. \cite{Alvey:2020xsk}.} DM \cite{Bar:2018acw}. Second, I will show how \emph{dynamical friction} can also allow us to study the properties of ultra-light fermionic DM.
Before moving to the main body of this work, I want to spend a few words on the landscape of DM models. A remarkable aspect of DM is that, despite the large amount of data pointing toward its existence \cite{Bertone:2010zza}, which furthermore comes from phenomena at very different time and length scales, not even the most fundamental property of DM (its mass) is known. If considered a particle, the constraints are roughly $m_{\rm DM}\gtrsim 10^{-22}$\,eV for bosonic fields \cite{Ferreira:2020fam} and $m_{\rm DM}\gtrsim 100$\,eV \cite{Tremaine:1979we}. Both of which can be derived from the extra forces that appear in dwarf spheroidals for the models at the limit\footnote{\label{foot}For the bosonic case, the occupation number is so large that a wave description is efficient, and backreacts when the distribution is compressed to distances below the de Broglie wavelength. More on this later. For the fermionic case, it is the degenerate pressure that backreacts if DM particles are compressed.}. Regarding the upper bound,  DM particles can be as heavy as allowed by the our effective description of Nature. It is customary to also mention the possibility of dark matter being made of macroscopic bodies of astrophysically significant mass \cite{Bertone:2010zza}.

\vspace{-.3cm}
\section{Galactic rotation curves to test ultra-light bosonic DM}
\label{sec:ULDM}

\vspace{-.3cm}
When shall we describe the light from a source as a classical field or as a collection of photons? The rough answer to this question is that the classical field emerges when the light configuration is made of several phase-space states occupied with large occupation numbers and uncorrelated phases \cite{Duncan:2012aja}. For dark matter in a galaxy (e.g. the Milky way) one can ask the same question, assuming it is bosonic. An order of magnitude estimate can be given as follows: the Milky Way halo is supposed to have a size of $\sim$100~kpc, and a maximum speed of $v\sim 2\times 10^{-3}$ for virialized matter (i.e. matter that remains bounded to the galaxy).
These numbers give a total number of states which can be filled in the relevant phase of the Milky Way of order $N_s\approx 10^{75}\left(m_{\rm DM}/\rm eV\right)^3$. The predicted mass of its DM halo being $M_{\rm DM}\approx 10^{12}\,M_{\odot}$ \cite{MW}, and assuming that DM is distributed universally over all the allowed states, 
on finds that this mass requires the occupation number 
\begin{equation}\label{eq:NP}
    N_p=M_{\rm DM}/(N_s m_{\rm DM})\approx 10^3 \left({\rm eV}/m_{\rm DM}\right)^4.
\end{equation} 
As a result, for $m_{\rm DM}\lesssim$\,eV, DM is better described using classical equations. In the simplest example of a scalar field $\phi$, the latter will correspond to the Klein-Gordon equation coupled to gravity. To understand the final configuration quantitatively, one can consider the typical wavelength of these waves $\lambda$. For scales larger than $\lambda$, the wavelike nature won't play any role, and these DM configurations can be described as a collection of free waves (or wavepackets) virializing through gravity and with typical velocity $\bar v$, corresponding to a typical wavelength $\bar \lambda\sim 1/(m_{\rm}\bar v)$. As a result, for distances $r\gg \bar \lambda$ (the `halo' part of the galaxy), the expected field is (up to normalization)
\begin{equation}
    \phi_{h}\sim \int_0^{v_{\rm max}} \mathrm{d}^3v e^{-v^2/\sigma_0^2} e^{i(\omega t-m \vec v \cdot \vec x)+\psi_v}+\mathrm{h.c.},\vspace{-.2cm}
    \end{equation}
where I included random phases $\psi_v$, a Maxwellian form for the distribution with $\sigma_0$ ($10^{-3}$ for the Milky Way) and $\omega=m\sqrt{1+v^2}$. From the previous formula, it is clear that most  modes (labelled by $v$) in this non-relativistic distribution,  oscillate with very similar frequencies, which can be considered coherent for times $t\lesssim 1/(m \sigma_0^2)$. Even if this is not part of this talk, I want to stress that these quasi-coherent oscillators generate \emph{unique} phenomena in the galactic halo, such as an extra source of gravitational heating
\cite{Marsh:2018zyw,Dalal:2022rmp,Bar-Or:2018pxz}. Closer to the galactic center, the backreaction from trying to compress waves below their de Broglie wavelength $\lambda_{dB}=/(mv)$ (sometimes called `quantum pressure') becomes efficient. This effect can be summarized by an extra term appearing in the Euler equation for the `fluid' degrees of freedom of the field configuration $\phi$, $\rho$ and $\vec v$ (see \cite{Ferreira:2020fam}), 
\begin{equation}
    \dot{\vec{v}}+\left(\vec{v} \cdot \frac{\nabla}{a}\right) \vec{v}=-\frac{\nabla}{a}\left(V-\frac{1}{2 m^2 a^2} \frac{\nabla^2 \sqrt{\rho}}{\sqrt{\rho}}\right), \label{eq:qp}
\end{equation}
where $V$ is the gravitational potential, and the last term is the one reacting when the field is contracted below $\lambda_{dB}\approx(10^{-22}{\rm eV}/m_{\rm DM})(10^{-3}/v)\,$kpc. Below these scales, numerical simulations\footnote{These simulations were also very relevant to show that the 
rest of the DM halo has a standard Navarro-Frenk-White (NFW) profile.} have shown the existence of a coherent structure at the center of galaxies, that have received the named of `soliton'  \cite{Schive:2014dra,Chan:2021bja}. Quite remarkably, one can understand the main features of these structures by considering a field configuration of the form $\phi\approx \frac{1}{\sqrt{2}m}e^{-i m_{\rm DM}t}e^{-i \gamma t}\chi(r)+\mathrm{h.c.},$, where we have assumed a spherically symmetric configuration and $\gamma\ll m$  is a parameter tat controls the non-relativistic limit. The final system of equations satisfied by $\chi(r)$ correspond to a self-gravitating configuration, and are not fixed by simply assuming that the field vanishes at large distances, see e.g. \cite{Bar:2018acw}. In fact, the scaling $\chi_\lambda(r)=\lambda^2 \chi_1(\lambda r)$ and $\gamma_\lambda=\lambda^2 \gamma$ generates a new solution. By fitting the configurations of the solitonic part of the halos in \cite{Schive:2014dra,Chan:2021bja}, one can see that they are fitted by the $\chi(r)$ functions found in the previous limit for a particular value of $\gamma$ for a given DM halo.  The numerical simulations yield a very intriguing/interesting relation that allows one to predict the size of the soliton for any DM halo up to some uncertainties. The first of these relations was noticed in \cite{Schive:2014hza}, and in \cite{Bar:2018acw} we expressed it as (this is valid up to factors of a few) 
\begin{equation}
\label{eq:EM}
    \left.\frac{E}{M}\right|_{\text {soliton}}\approx\left.\frac{E}{M}\right|_{\text {halo }},
\end{equation}
where $E/M$ is the total energy (including gravitational energy) divided by the mass of either the soliton or the halo.  This relation suggests some kind of equilibrium between the degrees of freedom of the halo and those of the soliton. The simulations performed in \cite{Chan:2021bja} found that a whole variety of relations is possible, and it is is still not clear if this corresponds to the fact that the configuration is not relaxed. In any case, what seems to be universal is the existence of the soliton in the center of galactic DM halos of considerable size and that changes the profile of DM at scales smaller than $\lambda_{dB}$, generating a core where otherwise one would expect a a cuspy NFW profile \cite{Bertone:2010zza}.

\begin{figure}[tbp]
\centering\vspace{-.9cm}
\includegraphics[width=.35\textwidth]{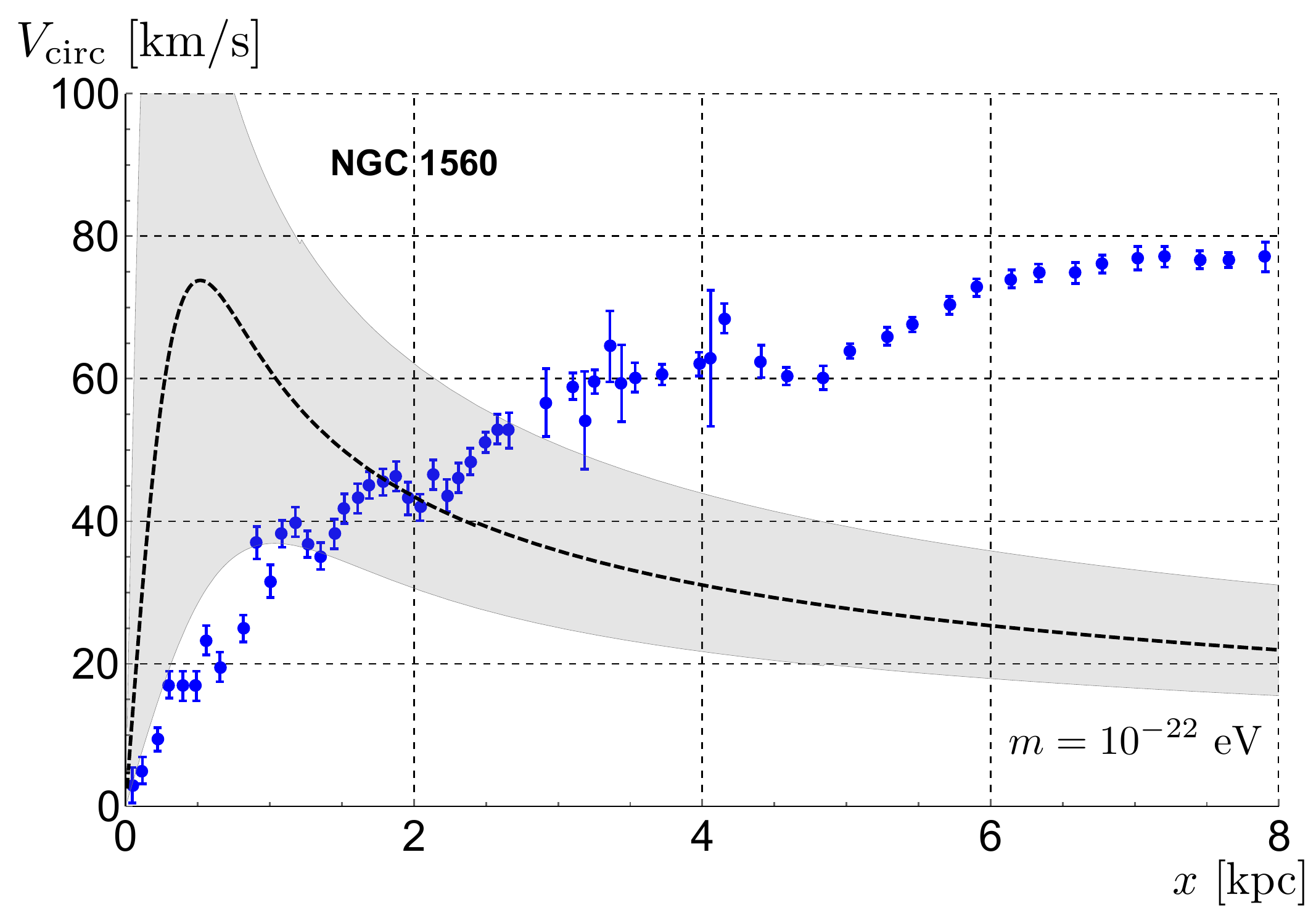}
~~
\includegraphics[width=.35\textwidth]{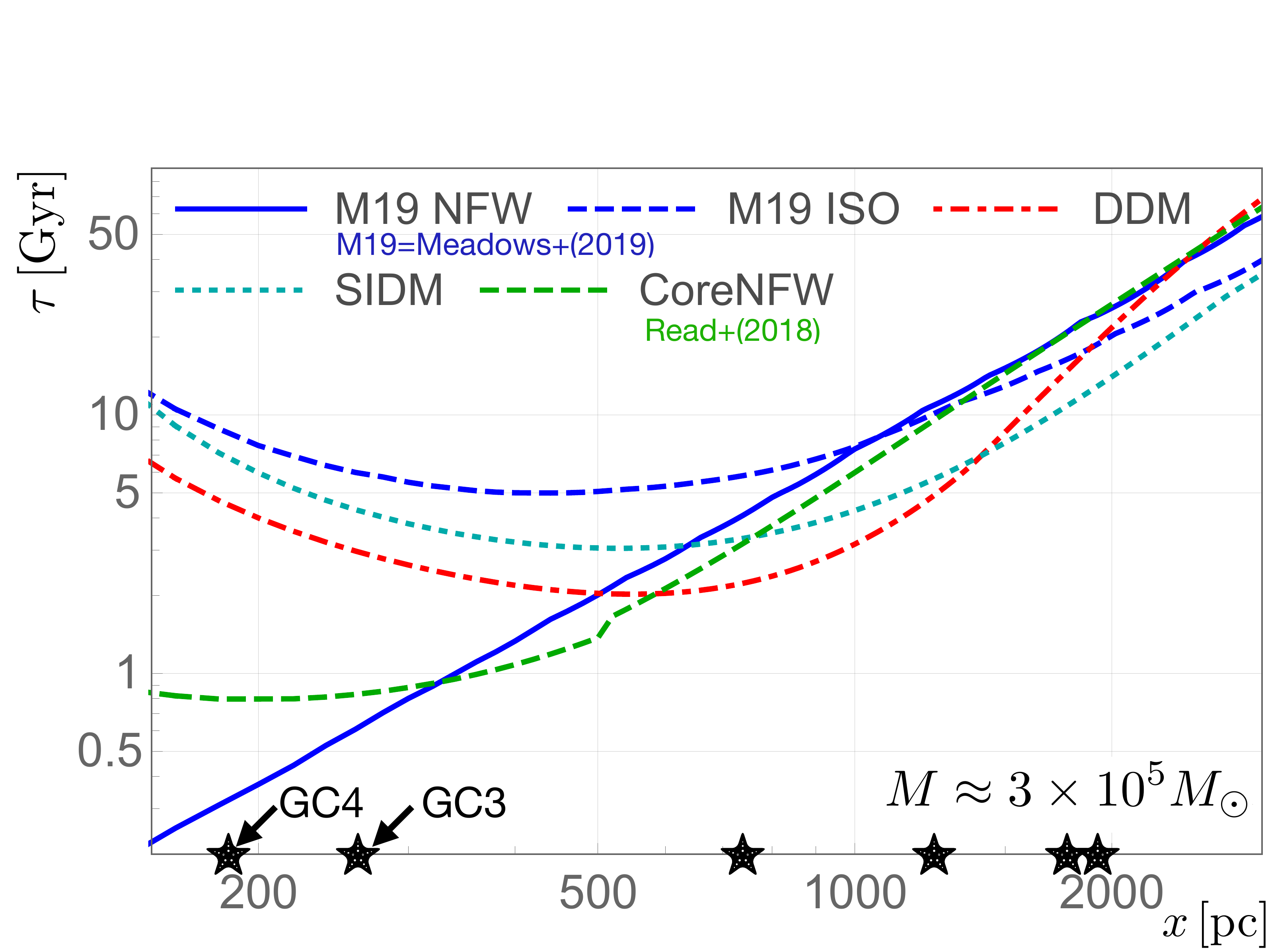}
\qquad
\caption{Left panel: rotation curves for NGC1560 vs the peak predicted by a soliton of $m_{\rm DM}\sim 10^{-22}$\,eV, from \cite{Bar:2018acw}. Right panel: typical time scale associated to dynamical friction for different models of DM, and GCs located at different radii $x$ of Fornax. One would like to have all these times above few Gyrs.}
\vspace{-.55cm}
\label{fig}
\end{figure}

If the DM profile is modified, its corresponding gravitational potential will also change. As a consequence, the properties of the visible matter connected to the average value of the DM potential (e.g. the rotation curves) may be modified. To model this, one can take a simplistic NFW profile at the outskirts of a galaxy, characterized by a concentration and a radius parameter, and use Eq.~\eqref{eq:EM} to find the corresponding soliton of this particular DM halo. The final gravitational potential will be a sum of $\Phi_{\rm halo}$ and $\Phi_{\rm soliton}$ and the expected orbital velocities will be modified as $V_{\text {circ }}^2=r \partial_r (\Phi_{\rm halo}+\Phi_{\rm soliton})$. A particularly interesting aspect is that the soliton profile generates a second peak in the velocity of rotation at small distances of the same size as the one generated by the halo part at large distances from the galactic center (see left panel of Fig.~\ref{fig}).  With this observation, one can now try to find the effect of the potential from the soliton into the rotation curves of galaxies. 
For this, in \cite{Bar:2018acw}, we used the SPARC sample \cite{Lelli:2016zqa}, which provides 
high resolution rotation curves of late-type  disc galaxy data. Furthermore, the  baryonic effects can also be estimated from the provided photometric data. As a result of not finding the inner peak associated to the soliton (see left panel of Fig.~\ref{fig}), or the corresponding changes in rotation curves produced by it (when the peak distance to the center is too small to be resolved), we concluded that it ULDM of masses below 
\begin{equation}
    m_{\rm DM}\lesssim 10^{-21}\, \rm eV  \label{eq:mdm}
    \vspace{-.3cm}
\end{equation}
are in tension with data. Since the size of the soliton scales as $1/m_{\rm DM}$, smaller masses are hard to probe, though, as also shown in \cite{Bar:2018acw}, our own galaxy may probe relevant in this case, since the gravitational potential of the Milky Way has been probed until $10^{-4}$pc. A couple of points are relevant: first, using a sample of hundreds of galaxies is very relevant to make sure that we are not looking at outlayers when searching for the soliton. Our method in \cite{Bar:2018acw} used  175 galaxies for this purpose. We also considered the possible influence of stars and  supermassive black holes in the centers of the galaxies, and concluded that our bound in \eqref{eq:mdm} is robust against it (see also \cite{2018MNRAS.478.2686C,Bar:2019bqz,Bar:2019pnz}).

\vspace{-.3cm}
\section{Dynamical friction to test light fermionic DM}
\vspace{-.3cm}

As I discussed above, the study of the velocity dispersion of dwarf spheroidals and their size, can be translated into an argument similar to the one behind \eqref{eq:NP} to claim that, because the occupation number of fermionic states cannot surpass 1, the bound $m\gtrsim 100$\,eV (known as Tremaine-Gunn bound) emerges \cite{Tremaine:1979we,Alvey:2020xsk}. A very important aspect that is not always emphasized is that, as the mass is reduced, and the phase space fills, it is not clear what the distribution should be. This is because it should correspond to a collection of (almost) degenerate fermions, out of equilibrium and interacting through gravity. These two points make numerical simulations quite challenging, as one should resort to quantum Vlasov equations, which I have not seen solved in the context of dark matter\footnote{A similar problem has been addressed in condensed matter. As I mentioned in my talk, this is a problem that interests me, if the reader wants to reach out to discuss it further.}. However, the relevance of this calculation seems limited, since the effects of degenerate DM as far as the distribution is concerned will be likely described by an effective temperature, related to the virial velocity, whose effect can be easily studied. 

The question we posed ourselved in \cite{Bar:2021jff} was what happens when you move \emph{across} this degenerate medium of fermions, and interact with it gravitationally. Would aspects of dynamical friction, tidal striping, or other dynamical processes where the DM halo interacts gravitationally (see \cite{Binney,2010gfe..book.....M}) be modified to observable levels? A similar question was asked in \cite{Hui:2016ltb} for ULDM, and several interesting directions were identified. In this contribution I only discuss dynamical friction (DF). The reason why we expect it to be modified in ULDM and light fermionic DM is the following. DF arises from the wake (a field overdensity) left behind as a perturber (e.g. a collection of stars, as a globular cluster) travels through a medium (e.g. the DM) with which it interacts gravitationally.
For ULDM, if this overdensity is smaller than $\lambda_{dB}$, the final term in \eqref{eq:qp}  responds to the contraction, and reduces the size of the wake, modifying DF. It turns out that the final answer is more complex, and still being investigated, se e.g \cite{Lancaster:2019mde,Bar-Or:2018pxz,Buehler:2022tmr,Wang:2021udl,Vicente:2022ivh}.
For the degenerate fermions, one expects that degenerate pressure  counteracts when one tries to compress it below the scale where fermions should occupy the same state (recall that for a galaxy, we assume a virial distribution, which as a related Fermi momentum). To realize this intuition, in \cite{Bar:2021jff} (see also \cite{Blas:2022ezo}) we followed a Fokker-Planck approach where a single perturber with momentum $p$ and distribution $f_1$ interacts with a medium with distribution $f_2$ through gravity. This  is encapsulated in a collision term
\begin{eqnarray}
    C\left[f_1\right]=\frac{(2 \pi)^4}{2 E_p} \int d \Pi_{k, p^{\prime}, k^{\prime}} \delta^{(4)}\left(p+k-p^{\prime}-k^{\prime}\right)|\overline{\mathcal{M}}|^2\left[f_1\left(p^{\prime}\right) f_2\left(k^{\prime}\right)\right. \nonumber
    \\
\left. ~~~   \left(1 \pm f_1(p)\right)\left(1 \pm f_2(k)\right)-f_1(p) f_2(k)\left(1 \pm f_1\left(p^{\prime}\right)\right)\left(1 \pm f_2\left(k^{\prime}\right)\right)\right].
\end{eqnarray}
In this formula, $E_p$ is the energy of the inital state,  $|\overline{\mathcal{M}}|$ represents the gravitational scattering cross-section from states with momentum $p$ (perturber) and $k$ (medium) to $p'$ and $k'$, and the fermionic/bosonic nature of the medium is represented by the relative signs in front of $f_2$. The final `friction' has the following effect in the velocity $\mathbf{V}$ of the perturber of mass $M$:
\begin{eqnarray}
    \frac{d \mathbf{V}}{d t}=-\frac{4 \pi G^2 \rho M \ln \Lambda}{V^3} F(V / \sigma) \mathbf{V},\label{eq:DFf}
\end{eqnarray}
where $\rho$ is the density of the medium, $M$ its mass, $G$ is the gravitational Newton constant, $\ln \Lambda$ represents a cut-off scale generating an $O(1)$ contribution and $F$ captures the details of the medium. For instance, for a Maxwellian distribution of particles with velocity dispersion $\sigma$  one gets $F\sim 1$ for $V\gg \sigma$ and $F\sim \frac{\sqrt{2}}{3 \sqrt{\pi}} \frac{V^3}{\sigma^3}$ for $V \ll \sigma$. For a degenerate dark matter (DDM) gas characterized by a Fermi velocity $v_F$, one finds that
\begin{equation}
   F_{\mathrm{DDM}}\rightarrow 1, \ {\rm for} \ V \gg v_F; ~~~ 
    F_{\mathrm{DDM}}\rightarrow \frac{V^3}{v_F^3}, \ {\rm for} \  V \ll v_F.
\vspace{-.2cm}
\end{equation}
This effect generates a reduction of dynamical friction for slow enough perturbers. Furthermore, the degenerate gas of fermions will generate a core structure that may reduce $\sigma$ and increase $\rho$. This core extends to distances $r_c \sim 681\left(\frac{\rho_0}{10^7 \mathrm{M}_{\odot} / \mathrm{kpc}^3}\right)^{-\frac{1}{6}}\left(\frac{g m^4}{2 \times(120 \mathrm{eV})^4}\right)^{-\frac{1}{3}} \mathrm{pc}$ (see also footnote~\ref{foot}).\\\vspace{-.2cm}

This last phenomena (modification of $\rho$ and/or $\sigma$) may also happen in ULDM, models with self-interacting dark matter (SIDM), or where baryonic feedback is modified. In \cite{Bar:2021jff} we considered different models  (SIDM, degenerate dark matter  with $m_{\mathrm{DM}} \approx 135\, \mathrm{eV}$, different baryonic feedback processes) that can fit the data connected to kinematics, and study if they can be differentiated with dynamical friction. For this, we focused on data from Fornax. This
very luminous nearby ($\sim 147$ kpc away) dwarf satellite has $\sim  4 \times 10^7 M_\odot$ stars in a $\sim$ kpc size. It is a dark matter dominated galaxy, with  5-6 globular clusters (GCs) with masses $\sim   10^5 M_\odot$, two of which at locations that seem very improbable, if dynamical friction has the size predicted by \eqref{eq:DFf} with $F\sim 1$. To understand this, one estimate the time scale for a perturber to plunging to the center of a galaxy (we assume $F\sim 1$)
\begin{equation}
    \tau \equiv \frac{|\mathbf{V}|}{|d \mathbf{V} / d t|} \sim 1.8\left(\frac{V}{12 \mathrm{~km} / \mathrm{s}}\right)^3 \frac{2 \times 10^7 \frac{M_{\odot}}{\mathrm{kpc}^3}}{\rho} \mathrm{Gyr},
    \end{equation}
    where we have included the values typical of Fornax.
This time scale is short enough to be suspicious of GCs at locations corresponding to time scales shorter than $\sim 10$\,Gpc.

By studying the data on the line-of-sight velocity dispersion of $\sim$ 2500 stars from Fornax, we are able to fit them with the aforementioned DM models (see \cite{Bar:2021jff,Blas:2022ezo}). More interestingly, these very same models produce \emph{significantly} different dynamical friction, since the medium properties are different. Still, all SIDM, more compact DM profiles and degenerate dark matter allows to increase $\tau$ to levels of $O(10)$ Gpc (the final answer being model dependent), and also agree with the kinematic properties of Fornax (see right panel of Fig.~\ref{fig}). As a result, one could invoke them to solve this `GC' Fornax timing problem\footnote{In \cite{Hui:2016ltb} a similar reasoning was used to suggest a mass of DM of $m_{\rm DM}\sim 10^{-22}$eV.}, though  it is hard to distinguish between them.

Before considering this possibility, one has to face a question about initial conditions of the GCs. Indeed, the probability to  end up with GCs at the positions we measure depends on the initial distribution. Unfortunately, the data from a single galaxy (Fornax) is not enough to make a statement, since the initial conditions are only known statistically.  Even assuming that Fornax is a good representive of the initial distribution of GCs, we showed that the standard DM effect does still produce GCs in the positions observed in Fornax with a 25\% probability. Hence, it is not clear to which extend Fornax poses a problem to standard DM models,  though the idea of using DF to differentiate among DM models is still sound. In particular, the most natural way forward is to  more data from systems where GCs or similar trackers are resolved in DM halos.
\vspace{-.2cm}

\section{Summary and outlook}

\vspace{-.3cm}
This contribution is devoted to new ideas to use galactic dynamics to differentiate between DM models. The final goal would be to implement the dynamics of different DM models into all the phenomenology related to the  DM halo. This should include rotation curves, but also many other dynamical phenomena, already known for  more standard DM models \cite{Bertone:2010zza,Binney,2010gfe..book.....M}.

%

I have focused on two particular examples. In Sec.~\ref{sec:ULDM}, I discussed some aspects in which the distribution of ULDM differs from that of standard DM. In particular, the existence of `quantum pressure' implies that  the field reacts when trying to  squeeze it below $\lambda_{dB}$. As a result, a coherent structure (soliton) appears in the center of DM halos. This soliton modifies the rotation curves to levels that seem to be  in tension with data for $m_{\rm DM}\lesssim 10^{-21}$eV. Higher masses requires measuring the properties of the galactic halo at distances below 0.1~kpc, and one could think about using the dynamical data from the Milky Way to explore them. This seems a good point to mention that other properties of ULDM may modify this conclusion,  e.g. self-interaction or the presence of supermassive black holes in the cases of higher masses. Also, it is important to recall that \cite{Chan:2021bja} found a condition between the solitons and the halo differing from \eqref{eq:EM}. Relevant for ULDM, the halo itself has patches with  wave behaviour, which modifies gravitational heating or dynamical friction. 

In the second part of this contribution I focused on degenerate DM fermions. For these, an optimistic bound of $m_{\rm DM}\gtrsim 100$\,eV arises from galactic dynamics. In the limiting case, small galaxies develop a core from the degenerate pressure. Furthermore, since the DM distribution is degenerate, one expects that it is harder to exchange momentum with it (since the `out-going' momentum state may be occupied). As a result, the effective coupling of a perturber with the DM halo should decrease, and enhance dynamical friction time scales. We showed in \cite{Bar:2021jff} that this intuition is correct, and that the time scales for perturbers that we see today in the DM halo to fall to the galactic center can be increased to levels of order $\sim$10 Gpc. However, the same conclusion can be achieved with other DM models. Furthermore, the mass considered $m_{\mathrm{DM}} \approx 135 \mathrm{eV}$, is almost excluded by galactic dynamics and hard to reconcile this small mass with other cosmological bounds \cite{Bar:2021jff,Carena:2021bqm}. As a possible future direction, I find it  interesting to understand which distribution is generated by a collection of degenerate fermions in `virial equilibrium' generated by gravity.

{\small
\textbf{Acknowledgements.} I thank the organizers of the 1st General Meeting and 1st Training School of the COST Action COSMIC WSIPers for their kind invitation. It is also a pleasure to thank all my great collaborators on the topics I discussed, N. Bar, K. Blum, H. Kim and S. Sibiryakov. I am also grateful to A. Caputo for comments on a previous draft. The research leading to these results has received
funding from the Spanish Ministry of Science and Innovation
(PID2020–115845GB-I00/AEI/10.13039/501100011033). IFAE is partially funded by the CERCA program of the Generalitat de Catalunya. I also acknowledge the support from the Departament de Recerca i Universitats de la Generalitat de Catalunya al Grup de Recerca i Universitats from Generalitat de Catalunya to the Grup de Recerca 00649 (Codi: 2021 SGR 00649).
}

\end{document}